\pdfoutput=1
\RequirePackage[l2tabu,orthodox]{nag}
\RequirePackage{fix-cm}
\documentclass{PoS}
\usepackage[T1]{fontenc}
\usepackage[utf8]{inputenc}
\usepackage{microtype}
\usepackage[artemisia]{textgreek}
\usepackage[Symbol]{upgreek}
\usepackage{mathtools}
\usepackage{csquotes}
\usepackage{physics}
\usepackage{booktabs}
\usepackage[nolist]{acronym}
\usepackage{siunitx}
\usepackage{tikz}
\usepackage{pgf}

\usepackage[style=base]{caption}
\usepackage{subcaption}
\NewDocumentCommand\cpi{}{\uppi}                        
\NewDocumentCommand\I{}{\mathrm{i}}
\NewDocumentCommand\e{ m }{\mathrm{e}^{#1}}
\NewDocumentCommand\Nf{}{N_\mathrm{f}}                        
\NewDocumentCommand\SU{}{\mathrm{SU}}
\NewDocumentCommand\Dhad{}{\Delta_{\mathrm{had}}}
\NewDocumentCommand\Dalphahad{}{\Delta\alpha_{\mathrm{had}}}
\NewDocumentCommand\thetaW{}{\theta_{\mathrm{W}}}
\NewDocumentCommand\sIIW{}{\sin^2\thetaW}
\NewDocumentCommand\sIIWh{}{\sin^2\hat{\theta}_{\mathrm{W}}}
\NewDocumentCommand\SPi{}{\bar{\Pi}}
\NewDocumentCommand\tauint{}{\tau_{\mathrm{int}}}
\NewDocumentCommand\tcut{}{x_0^{\mathrm{cut}}}
\NewDocumentCommand\ti{}{x_{0i}}
\NewDocumentCommand\tnotsym{}{t_0^{\mathrm{sym}}}
\makeatletter
\newcommand*{\textoverline}[1]{$\overline{\hbox{#1}}\m@th$}
\makeatother
\NewDocumentCommand\MSbar{}{\textoverline{\scshape ms}}
\DeclareSIUnit\fm{\femto\metre}

\begin{acronym}
  \acro{SM}{Standard Model}
  \acro{QFT}{quantum field theory}
  \acro{QCD}{quantum chromodynamics}
  \acro{HVP}{hadronic vacuum polarization}
  \acro{TMR}{time-momentum representation}
  \acro{CCS}{Lorentz-covariant coordinate space}
  \acro{FFT}{fast Fourier transform}
  \acro{CLS}{Coordinated Lattice Simulations}
  \acro{BC}{boundary condition}
  \acro{OBC}{open boundary condition}
  \acro{GS}{Gounaris-Sakurai}
  \acro{SN}[S/N]{signal-to-noise ratio}
\end{acronym}

\let\OLDthebibliography\thebibliography
\renewcommand\thebibliography[1]{
  \OLDthebibliography{#1}
  \setlength{\parskip}{0pt}
  \setlength{\itemsep}{0pt plus 0.3ex}
}

\title{The hadronic contribution to the running\\ of the electromagnetic coupling\\ and the electroweak mixing angle}

\ShortTitle{The hadronic running of the electroweak couplings}

\author{Marco Cè,\thanks{Speakers.} $^{ab}$\footnote{Current affiliation: Theoretical Physics department, CERN, Geneva, Switzerland}\ \ %
        Teseo San José,\addtocounter{footnote}{-2}\footnotemark\ $^{abc}$ Antoine Gérardin,$^d$ Harvey B.\ Meyer,$^{abc}$ Kohtaroh Miura,$^{abe}$ Konstantin Ottnad,$^{bc}$ Andreas Risch,$^{bc}$ Jonas Wilhelm,$^{bc}$ Hartmut Wittig$^{abc}$\\
        \llap{$^a$}Helmholtz-Institut Mainz, Johannes Gutenberg-Universität Mainz, Germany\\
        \llap{$^b$}PRISMA\textsuperscript{+} Cluster of Excellence, Johannes Gutenberg-Universität Mainz, Germany\\
        \llap{$^c$}Institut für Kernphysik, Johannes Gutenberg-Universität Mainz, Germany\\
        \llap{$^d$}John von Neumann-Institut für Computing (NIC), DESY Zeuthen, Germany\\
        \llap{$^e$}Kobayashi-Maskawa Institute for the Origin of Particles and the Universe, Nagoya University, Japan\\
        E-mail: \email{marco.ce@uni-mainz.de}, \email{msanjosp@uni-mainz.de}
}
\makeatletter
	\global\@speakertrue
	\global\setbox\@firstaubox
	\hbox{{\let\thanks\@gobble
		\let\footnote\@gobble\small
		\rm Marco Cè and Teseo San José}}%
\makeatother

\abstract{
  The electromagnetic coupling $\alpha$ and the electroweak mixing angle $\thetaW$ are parameters of the \ac{SM} that enter precision \ac{SM} tests and play a fundamental rôle in beyond \ac{SM} physics searches.
  Their values are energy dependent, and non-perturbative hadronic contributions are the main source of uncertainty to the theoretical knowledge of the running with energy.
  We present a lattice study of the leading hadronic contribution to the running of $\alpha$ and $\sIIW$.
  The former is related to the \ac{HVP} function of electromagnetic currents, and the latter to the \ac{HVP} mixing of the electromagnetic current with the vector part of the weak neutral currents.
  We use the \ac{TMR} method to compute the \ac{HVP} on the lattice, estimating both connected and disconnected contributions on $\Nf=2+1$ non-perturbatively $\order{a}$-improved Wilson fermions ensembles from the \ac{CLS} initiative.
  The use of different lattice spacings and quark masses allows us to reliably extrapolate the results to the physical point.
\acresetall\vspace*{0.5cm}
\begin{flushright}
  MITP/19-064\\
  DESY 19-179
\end{flushright}
}

\FullConference{37th International Symposium on Lattice Field Theory - Lattice2019\\
		16-22 June 2019\\
		Wuhan, China}

\begin{document}
\acused{QCD}
\section{The running of the electromagnetic coupling}

The predictions of the \ac{SM} of particle physics are today tested to a high degree of precision in experiments that span a vast range of energy scales, from atomic physics to high-energy colliders.
The connection between these energy scales is encoded in the \emph{running} with energy of the strength of interactions.
For instance, the fine-structure constant $\alpha=1/\num{137.035999139(31)}$~\cite{Tanabashi:2018oca} is known from low-energy experiments to better than a part per billion.
However, its effective value for physics around the $Z$ pole is $\hat{\alpha}^{(5)}(M_Z)=1/\num{127.955(10)}$~\cite{Tanabashi:2018oca}, a \SI{7}{\percent} larger value.
In the \emph{on-shell} scheme, the running of $\alpha$ at a given time-like momentum transfer $q^2$ is described by
\begin{equation}
  \alpha(q^2) = \frac{\alpha}{1-\Delta\alpha(q^2)} ,
\end{equation}
in terms of the $\Delta\alpha(q^2)$ function.
While the lepton contribution to $\Delta\alpha(q^2)$ can be computed in perturbation theory, the estimate of the quark contribution at low energies requires non-perturbative calculations of hadronic physics.
Conventionally, the hadronic contribution $\Dalphahad(q^2)$ is related through the optical theorem to the $R$-ratio, \emph{i.e.}\ the total cross-section $e^+e^-\to\text{hadrons}$ over $e^+e^-\to\mu^+\mu^-$, which is estimated using a compilation of experimental data.
This results in $\Dalphahad^{(5)}(M_Z^2)=\num{0.02764(7)}$~\cite{Tanabashi:2018oca} constituting the main contribution to the uncertainty on $\alpha(M_Z^2)$.
In an alternative approach~\cite{Jegerlehner:2011mw}, the Adler function $D(Q^2)$ at space-like $Q^2=-q^2$ is estimated from the same data and used to compute the running up to $Q^2\approx\SI{2}{\GeV\squared}$.
Around this $Q^2$ and above, $D(Q^2)$ can be computed reliably in perturbative \ac{QCD}.
Crucially, $\Dalphahad(-Q^2)$ at space-like momenta $Q^2>0$ is accessible to non-perturbative lattice techniques defined in Euclidean space-time~\cite{Burger:2015lqa,Francis:2015grz,Borsanyi:2017zdw}.
It is given by
\begin{equation}
\label{eq:Dhad_alpha}
  \Dalphahad(-Q^2) = 4\cpi\alpha \SPi^{\gamma\gamma}(Q^2) , \qquad \SPi^{\gamma\gamma}(Q^2) = [\Pi^{\gamma\gamma}(Q^2)-\Pi^{\gamma\gamma}(0)] ,
\end{equation}
where $\SPi(Q^2)$ is the subtracted \ac{HVP} function\footnote{
  In this work we denote with $\SPi(Q^2)$ with $Q^2>0$ the \ac{HVP} function at \emph{space-like} momenta.
}
\begin{equation}
  (Q_\mu Q_\nu-\delta_{\mu\nu}Q^2)\Pi^{\gamma\gamma}(Q^2) = \Pi_{\mu\nu}^{\gamma\gamma}(Q^2) = \int\dd[4]{x} \e{\I Q\cdot x} \ev{j_\mu^\gamma(x) j_\nu^\gamma(0)}
\end{equation}
of the electromagnetic current
\begin{equation}
  j_\mu^\gamma = \frac{2}{3}\bar{u}\gamma_\mu u - \frac{1}{3}\bar{d}\gamma_\mu d - \frac{1}{3}\bar{s}\gamma_\mu s + \frac{2}{3}\bar{c}\gamma_\mu c .
\end{equation}

Lattice \ac{QCD} can thus provide an estimate that does not depend on experimental $R$-ratio data and cross-check the phenomenological estimate.
This is of great interest in the context of global \ac{SM} fits, where $\Dalphahad(M_Z^2)$ is an input.
Indeed, the best-fit result for the Higgs mass excluding kinematic constraints is $M_H=\SI[parse-numbers=false]{90^{+17}_{-16}}{\GeV}$, $1.9\sigma$ below the measured value, and a shift of $\pm\num{e-4}$ in $\Dalphahad^{(3)}(\SI{4}{\GeV\squared})$ corresponds to a shift of $\mp\SI{4.5}{\GeV}$ in $M_H$~\cite{Tanabashi:2018oca}.
This is connected to the tension between the \ac{SM} and experimental determinations of the anomalous magnetic moment of the muon $(g-2)_\mu$ (see Ref.~\cite{Meyer:2018til} for a review), because a solution of the $(g-2)_\mu$ puzzle that involves an increase of the \ac{SM} estimate of the leading hadronic contribution $a_\mu^{\mathrm{HLO}}$ has to avoid a correlated increase of $\Dalphahad$~\cite{Passera:2008jk}.

A second connection to $(g-2)_\mu$ comes from the high-precision measurement of $\alpha(-Q^2)$ at space-like $Q^2$ in $t$-channel scattering proposed by the MUonE collaboration~\cite{Abbiendi:2016xup,Abbiendi:2677471}.
Isolating the hadronic contribution $\Dalphahad(-Q^2)$, an independent determination of $a_\mu^{\mathrm{HLO}}$ is obtained from~\cite{Lautrup:1971jf}
\begin{equation}
  a_\mu^{\mathrm{HLO}} = \frac{\alpha}{\cpi} \int_0^1 \dd{x} (1-x) \Delta\alpha_{\mathrm{had}}(t(x)), \qquad t(x) = \frac{x^2m_\mu^2}{x-1} \leq 0 .
\end{equation}
The proposed experiment is limited to $x<0.932$, corresponding to $Q^2\lesssim\SI{0.14}{\GeV\squared}$, that leaves out \SI{13}{\percent} of the $a_\mu^{\mathrm{HLO}}$ integral.
Lattice data can complement the experiment for $Q^2\gtrsim\SI{0.14}{\GeV\squared}$~\cite{Marinkovic:Lattice2018}.

\section{The running of the electroweak mixing angle}

As a second quantity, we consider the electroweak mixing angle or Weinberg angle $\thetaW$, that is, the parameter of the Standard Model of particle physics that parametrizes the mixing between electromagnetic and weak interactions
\begin{equation}
\label{eq:Dhad_sIIW}
  \sIIW = \frac{g'^2}{g^2+g'^2} , \qquad e = g \sin\thetaW = g' \cos\thetaW ,
\end{equation}
where $g$ and $g'$ are the $\mathrm{SU}(2)_L$ and $\mathrm{U}(1)_Y$ couplings, respectively.
Beyond tree level, as the couplings themselves, its precise value is scheme and energy dependent.
In a given scheme, the mixing angle is measured to sub-permille precision at energies around $M_Z$, \emph{e.g.}\ \emph{on-shell} $\sIIW=1-M_W^2/M_Z^2=\num{0.22343(7)}$ or in the \MSbar-scheme $\sIIWh(M_Z)=\num{0.23122(7)}$~\cite{Tanabashi:2018oca}.
Conversely, the mixing angle in the Thomson limit, $\abs*{q^2}\ll m_e^2$, can be defined in a scheme-independent way~\cite{Ferroglia:2003wa,Kumar:2013yoa}.
Its experimental value is less well known, but upcoming experiments target up to a \SI{0.15}{\percent} precision at a momentum transfer of \SI{4.5e-3}{\GeV\squared}~\cite{Benesch:2014bas,Becker:2018ggl}.
A more precise value, $\sIIWh(0)=\num[parse-numbers=false]{0.238\,68(5)(2)}$~\cite{Erler:2017knj}, is obtained computing the running in the \MSbar-scheme from $M_Z$ to low energies.
The first error is the uncertainty on the $Z$-pole value, while the second error is the total theoretical uncertainty on the running and it is dominated by the non-perturbative hadronic contribution.

The energy dependence of $\sIIW$ in the on-shell scheme can be written as
\begin{equation}
  \sIIW(q^2) = \sIIW \left[ 1 + \Delta\sIIW(q^2) \right] ,
\end{equation}
where $\sIIW$ is the value in the low-energy limit.
Similarly to Eq~\eqref{eq:Dhad_alpha}, the leading hadronic contribution to the running at space-like $Q^2=-q^2$ is given by~\cite{Jegerlehner:1985gq,Jegerlehner:2011mw}
\begin{equation}
  \Dhad\sin^2 \thetaW(-Q^2) = -\frac{e^2}{\sIIW} \SPi^{Z\gamma}(Q^2) ,
\end{equation}
where $\SPi^{Z\gamma}(Q^2)$ is \ac{HVP} mixing of the electromagnetic current $j_\mu^\gamma$ and the vector part of the neutral weak current $j_\mu^Z$
\begin{equation}
  j_\mu^Z \big|_\mathrm{vector}      = j_\mu^{T_3} \big|_\mathrm{vector} - \sIIW j_\mu^\gamma , \qquad
  j_\mu^{T_3} \big|_\mathrm{vector}  = \frac{1}{4}\bar{u}\gamma_\mu u - \frac{1}{4}\bar{d}\gamma_\mu d - \frac{1}{4}\bar{s}\gamma_\mu s + \frac{1}{4}\bar{c}\gamma_\mu c .
\end{equation}
As in the electromagnetic case, $\SPi^{Z\gamma}(Q^2)$ is directly accessible to lattice computations~\cite{Burger:2015lqa,Francis:2015grz,Guelpers:2015nfb,Ce:2018ziv}.

\section{The \acs{TMR} method}
\label{sec:TMR}

\begin{figure}[t]
  \centering
  \scalebox{.62}{\input{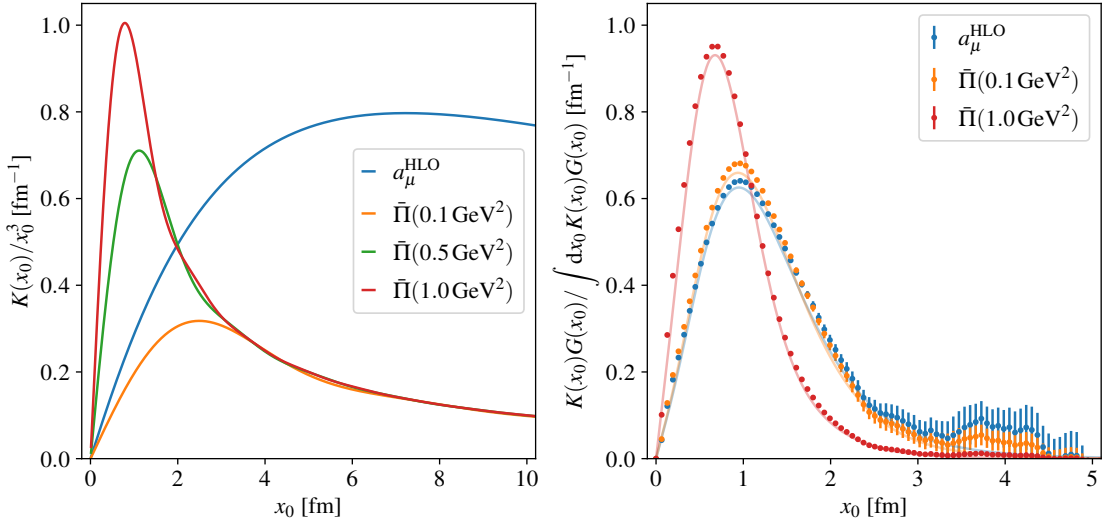}}
  \caption{%
    Left: the kernel $K(x_0,Q^2)$ of the \acs{TMR} integral in Eq.~\eqref{eq:TMRmethod} divided by $x_0^3$ for different values of $Q^2$, compared to the kernel for $a_\mu^{\mathrm{HLO}}$~\cite{Bernecker:2011gh,DellaMorte:2017dyu} (blue line), as a function of time $x_0$.
    Right: contribution of $G(x_0)K(x_0,Q^2)$ to the \acs{TMR} integral normalized to the value of the integral, comparing different kernels $K(x_0)$.
    The light coloured lines are drawn using a model for the Euclidean-time correlator $G(x_0)$~\cite{Bernecker:2011gh}, that is also used for the integral, while the data points with error bars are obtained using actual lattice correlator data at the physical pion mass.
  }\label{fig:kernel}
\end{figure}

The computation of $\SPi^{\gamma\gamma}$ and $\SPi^{Z\gamma}$ as functions of $Q^2$ is similar to that of $a_\mu^{\mathrm{HLO}}$, the leading-order \ac{HVP} contribution to $(g-2)_\mu$, and, as in that case, different methods are available, such as the four-momentum \emph{hybrid method}, the time moments, or the \ac{TMR} method~\cite{Bernecker:2011gh,Francis:2013qna}.
In this study, we employ the \ac{TMR} method to compute the subtracted \ac{HVP} function
\begin{equation}
\label{eq:TMRmethod}
  \SPi(Q^2) = \int_0^\infty \dd{x_0} G(x_0) \left[ x_0^2 - \frac{4}{Q^2}\sin[2](\frac{Qx_0}{2})\right] , \qquad
  G(x_0) = -\frac{1}{3} \int\dd[3]{x} \sum_{k=1}^3 \ev{ j_k(x) j_k(0) } ,
\end{equation}
where we have to integrate over Euclidean time the product of the zero-momentum-projected correlator, $G^{\gamma\gamma}(x_0)$ or $G^{Z\gamma}(x_0)$, times a $Q^2$-dependent kernel $K(x_0,Q^2)=x_0^2-(4/Q^2)\sin^2(Qx_0/2)$.
This allows us, in principle, to input any value of $Q^2$ in the kernel.
The properties of the kernel significantly influence the systematics of the integral: On the one hand, a shorter-range kernel puts a larger weight on the correlator at short times.
Since the correlator on the lattice is sampled at a spacing $a$, one needs $Q^2\ll(\cpi/a)^2$ in order to avoid large cut-off effects.
On the other hand, a longer-range kernel weights relatively more the long-time behaviour of the correlator, which is noisier and susceptible to finite-volume effects.
Different kernels are compared in the left plot of Figure~\ref{fig:kernel}, including the one used to compute $a_\mu^{\mathrm{HLO}}$, Eq.~(84) from Ref.~\cite{Bernecker:2011gh}.
In the right plot we show the corresponding relative contribution to the integral against the time variable.
The \ac{HVP} function $\SPi$ at $Q^2=\SI{0.1}{\GeV\squared}$ receives the larger contribution around \SI{1}{\fm}, as $a_\mu^{\mathrm{HLO}}$ does, but at long times the contribution to the former is smaller.
When lattice data at physical pion masses is used for $G(x_0)$, this results in a smaller statistical error on $\SPi(\SI{0.1}{\GeV\squared})$ than on $a_\mu^{\mathrm{HLO}}$.
In the case of $\SPi(\SI{1}{\GeV\squared})$, the correlator tail has a negligible contribution to the statistical error.

\section{Lattice setup}

We perform the computation on the $\Nf=2+1$ set of ensembles from the \ac{CLS} initiative~\cite{Bruno:2014jqa}, with tree-level Lüscher-Weisz gauge action and non-perturbatively $\order{a}$-improved Wilson fermions.
The list of ensembles employed in this work is in Table~\ref{tab:ensemble}.
We use four lattice spacings,
$u$ and $d$ quark masses are degenerate, thus we have exact isospin symmetry, and the pseudoscalar meson masses span from $M_\pi=M_K\approx\SI{415}{\MeV}$ at the $\SU(3)$-symmetric point to the physical ones along a trajectory on which the sum of the bare $u$, $d$ and $s$ quark masses is kept constant.
We set the scale using $(8t_0^{\mathrm{phys}})^{1/2}=\SI[parse-numbers=false]{0.415(4)(2)}{\fm}$~\cite{Luscher:2010iy,Bruno:2016plf}.

\begin{table}[tb]
  \caption{%
    List of ensembles from the \acs{CLS} initiative employed, with approximate lattice spacings, sizes and pion and kaon masses.
    A * denotes the ensembles for which the disconnected contribution is available.
    All ensembles have open boundary conditions in time, except those denoted with \textsuperscript{§} that have periodic boundary conditions in time.
    Values of $\tnotsym$ and $a$ are from Ref.~\cite{Bruno:2016plf}.
  }\label{tab:ensemble}
  \centering
  \begin{tabular}{lS[table-format=3]S[table-format=2]S[table-format=1.3]S[table-format=1.3]S[table-format=1.1]S[table-format=3]S[table-format=3]S[table-format=1.1]}
    \toprule
    & {$T/a$} & {$L/a$} & {$\tnotsym/a^2$} & {$a$ [\si{\fm}]} & {$L$ [\si{\fm}]} & \multicolumn{2}{c}{$M_\pi$, $M_K$ [\si{\MeV}]} & {$M_\pi L$} \\
    \midrule
    H101  &  96 & 32 & 2.860 & 0.086 & 2.8 & \multicolumn{2}{c}{$415$} & 5.8 \\
    H102  &  96 & 32 &       &       & 2.8 & 355 & 440 & 5.0 \\
    H105* &  96 & 32 &       &       & 2.8 & 280 & 460 & 3.9 \\
    N101  & 128 & 48 &       &       & 4.1 & 280 & 460 & 5.8 \\
    C101* &  96 & 48 &       &       & 4.1 & 220 & 470 & 4.6 \\
    \midrule
    B450\textsuperscript{§} &  64 & 32 & 3.659 &  0.076 & 2.4 & \multicolumn{2}{c}{$415$} & 5.1 \\
    S400  & 128 & 32 &       &        & 2.4 & 350 & 440 & 4.3 \\
    N401* & 128 & 48 &       &        & 3.7 & 285 & 460 & 5.3 \\
    \midrule
    H200  &  96 & 32 & 5.164 & 0.064 & 2.1 & \multicolumn{2}{c}{$420$} & 4.4 \\
    N202  & 128 & 48 &       &       & 3.1 & \multicolumn{2}{c}{$410$} & 6.4 \\
    N203* & 128 & 48 &       &       & 3.1 & 345 & 440 & 5.4 \\
    N200* & 128 & 48 &       &       & 3.1 & 285 & 465 & 4.4 \\
    D200* & 128 & 64 &       &       & 4.1 & 200 & 480 & 4.2 \\
    E250*\textsuperscript{§} & 192 & 96 &       &      & 6.2 & 130 & 490 & 4.1 \\
    \midrule
    N300  & 128 & 48 & 8.595 & 0.050 & 2.4 & \multicolumn{2}{c}{$420$} & 5.1 \\
    N302* & 128 & 48 &       &       & 2.4 & 345 & 460 & 4.2 \\
    J303  & 192 & 64 &       &       & 3.2 & 260 & 475 & 4.2 \\
    \bottomrule
  \end{tabular}
\end{table}

\subsection{Flavour decomposition and renormalization}

The correlators are computed on the ensembles in Table~\ref{tab:ensemble} as described in Ref.~\cite{Gerardin:2019rua}, to which we refer to for the unexplained notation.
At the sink, we employ both the local and conserved discretizations of the vector current
\begin{subequations}
\begin{gather}
  j_\mu^\mathrm{l}(x) = \bar{q}(x) \gamma_\mu q(x) , \\
  j_\mu^\mathrm{c}(x) = \frac{1}{2}\left[ \bar{q}(x+a\hat{\mu})(1+\gamma_\mu)U^\dagger_\mu(x)q(x) - \bar{q}(x)(1-\gamma_\mu)U_\mu(x)q(x+a\hat{\mu}) \right] ,
\end{gather}
\end{subequations}
while only the local current is used at the source.
The currents are non-perturbatively $\order{a}$-improved and renormalized.
To this purpose, we introduce a flavour $\SU(3)$ decomposition of the current.
For the local discretization, we have~\cite{Bhattacharya:2005rb}
\begin{subequations}
\begin{gather}
  j_{\mu,R}^{3,\mathrm{l}} = Z_V \left(1 + 3\bar{b}_Vam_q^{\mathrm{av}} + b_Vam_{q,\ell}\right) j_\mu^{3,\mathrm{il}} , \\
  \begin{pmatrix}
    j_\mu^8 \\
    j_\mu^0
  \end{pmatrix}_R^{\mathrm{l}} = Z_V \begin{pmatrix}
    1 + 3\bar{b}_Vam_q^{\mathrm{av}} + b_V\frac{a(m_{q,\ell}+2m_{q,s})}{3} & \left(\frac{b_V}{3}+f_V\right)\frac{2a(m_{q,\ell}-m_{q,s})}{\sqrt{3}} \\
    r_V d_V\frac{a(m_{q,\ell}-m_{q,s})}{\sqrt{3}} & r_V + r_V (3\bar{d}_V+d_V)am_q^{\mathrm{av}}
  \end{pmatrix} \begin{pmatrix}
    j_\mu^8 \\
    j_\mu^0
  \end{pmatrix}^{\mathrm{il}} , 
\end{gather}
\end{subequations}
where the improved non-singlet and singlet local currents are
\begin{equation}
  j_\mu^{a,\mathrm{il}} = j_\mu^{a,\mathrm{l}} + ac_V^{\mathrm{l}}\tilde{\partial}_\nu T_{\nu\mu}^a , \qquad j_\mu^{0,\mathrm{il}} = j_\mu^{0,\mathrm{l}} + a\bar{c}_V^{\mathrm{l}}\tilde{\partial}_\nu T_{\nu\mu}^0 ,
\end{equation}
and the breaking of flavour $\SU(3)$ symmetry introduces a mixing between the singlet and non-singlet $I=0$ components.
For the conserved discretization, no renormalization is needed
\begin{equation}
  j_{\mu,R}^{a,\mathrm{c}} = j_\mu^{a,\mathrm{c}} + ac_V^{\mathrm{c}}\tilde{\partial}_\nu T_{\nu\mu}^a , \qquad j_{\mu,R}^{0,\mathrm{c}} = j_\mu^{0,\mathrm{c}} + a\bar{c}_V^{\mathrm{c}}\tilde{\partial}_\nu T_{\nu\mu}^0 .
\end{equation}
We use the renormalization and improvement coefficients determined non-perturbatively in Ref.~\cite{Gerardin:2018kpy}.
Since the coefficients to renormalize the singlet local current are unknown, we use only the conserved vector current for the singlet component.
This implies that we can only put the $Z$ current at the sink, where the conserved discretization is available.
Moreover, we set $f_V=0$ and $\bar{c}_V^{\mathrm{c},\mathrm{l}}=c_V^{\mathrm{c},\mathrm{l}}$, which is valid up to $\order{g_0^5}$ and introduces a negligible error.

The charm contribution is also computed in the quenched approximation, with the charm quark mass tuned using the experimental $D_s$ meson mass and the local current renormalized computing the mass-dependent $Z_V^c$~\cite{Gerardin:2019rua}.
Including the latter contribution, the $\gamma\gamma$ and $Z\gamma$ bare correlators are
\begin{subequations}
\label{eq:corr_flavour}
\begin{gather}
  G_{\mu\nu}^{\gamma\gamma}(x) = G_{\mu\nu}^{33}(x) + \frac{1}{3}G_{\mu\nu}^{88}(x) + \frac{4}{9} C_{\mu\nu}^{c,c}(x), \\
  G_{\mu\nu}^{Z\gamma}(x) =  \left(\frac{1}{2}-\sIIW\right)G_{\mu\nu}^{\gamma\gamma}(x) - \frac{1}{6\sqrt{3}} G_{\mu\nu}^{08}(x) - \frac{1}{18} C_{\mu\nu}^{c,c}(x) ,
\end{gather}
\end{subequations}
where the flavour $\SU(3)$ contributions are defined as\footnote{%
  In the usual lattice notation, $G_{\mathrm{conn}}^\ell=2G^{33}$ and $G_{\mathrm{conn}}^s=3G_{\mathrm{conn}}^{88}-G^{33}$.
  Moreover, $G^{08}_{\mathrm{conn}}=\sqrt{3}(G^{33}-G_{\mathrm{conn}}^{88})/2$.
}
\begin{subequations}
\begin{gather}
  G_{\mu\nu}^{33}(x) = \frac{1}{2}C^{\ell,\ell}_{\mu\nu}(x) , \\
  G_{\mu\nu}^{88}(x) = \frac{1}{6}\left[ C^{\ell,\ell}_{\mu\nu}(x) + 2C^{s,s}_{\mu\nu}(x) + 2D^{\ell-s,\ell-s}_{\mu\nu}(x) \right] , \\
  G_{\mu\nu}^{08}(x) = \frac{1}{2\sqrt{3}}\left[ C^{\ell,\ell}_{\mu\nu}(x) - C^{s,s}_{\mu\nu}(x) + D^{2\ell+s,\ell-s}_{\mu\nu}(x) \right] ,
\end{gather}
\end{subequations}
and the connected and disconnected Wick's contractions are
\begin{subequations}
\begin{gather}
  C^{f_1,f_2}_{\mu\nu}(x) = -\ev{\Tr{D_{f_1}^{-1}(x,0)\gamma_\mu     D_{f_2}^{-1}(0,x)\gamma_\nu}} ,\\
  D^{f_1,f_2}_{\mu\nu}(x) =  \ev{\Tr{D_{f_1}^{-1}(x,x)\gamma_\mu}\Tr{D_{f_2}^{-1}(0,0)\gamma_\nu}} .
\end{gather}
\end{subequations}

\subsection{Autocorrelation study}
\enlargethispage{1\baselineskip}

Due to the update procedure of Monte Carlo simulations different configurations within one chain are not independent.
In order to give a reliable estimate of the observable's uncertainty we need to take these autocorrelations into account.
For this work we have used the $\Gamma$-method~\cite{Wolff:2003sm,DePalma:2017lww} to estimate the autocorrelation time, $\tauint$ and, from it, an appropriate bin size to obtain statistically independent samples.
Since autocorrelations are observable-dependent we have computed $\tauint$ of $\Dalphahad$ for the light and strange flavours on each ensemble at different energies.
Extra caution needs to be taken regarding the tail of the correlator, which shows the \acl{SN} problem (see Section~\ref{sec:SN}).
To obtain a good estimate of $\tauint$ we only take into account the correlator until a certain maximum value of $x_0$.
Repeating the process for different time cuts, we know where the noise of the tail starts to dominate and the value of $\tauint$.
We choose a bin size $B=2\tauint$ for each ensemble to bin the correlator data and obtain a distribution of independent bootstrap samples to carry out the main analysis.


\subsection{Finite-size corrections}

In this work, we estimate finite-volume corrections using the same strategy described in Ref.~\cite{Gerardin:2019rua}.
Namely, we compute the difference between the infinite- and finite-volume $I=1$ correlator and apply it as a correction to the lattice data.
We use two distinct models to compute the difference, depending on whether the correlator is considered at short or long times.

At long times, we use the $I=1$ correlator obtained from the time-like pion form factor $F_\pi(\omega)$.
In infinite volume, that is
\begin{equation}
  G^{33}(x_0,\infty) = \int_0^\infty \dd{\omega}\omega^2 \rho(\omega^2) \e{-\omega x_0} , \qquad
  \rho(\omega^2) = \frac{1}{48\cpi^2} \left( 1-\frac{4M_\pi^2}{\omega^2} \right)^{\frac{3}{2}} \abs{F_\pi(\omega)}^2 ,
\end{equation}
while the finite volume expression is computed from the Lüscher energies $\omega_n$ and the Lellouch-Lüscher amplitudes,
\begin{equation}
  G^{33}(x_0,L) = \sum_n \abs{A_n}^2 \e{-\omega_n x_0} ,
\end{equation}
extracted applying the Lellouch-Lüscher formalism~\cite{Luscher:1991cf,Lellouch:2000pv,Meyer:2011um}.
In this work, we use the \ac{GS} parametrization~\cite{Gounaris:1968mw} of $F_\pi(\omega)$, that depends on two parameters, $m_\rho$ and $g_{\rho\pi\pi}$, which are obtained either by fitting the correlator at long times or by a spectroscopic analysis~\cite{Gerardin:2019rua}.
We emphasise that the model is used only to correct for the relatively small finite-volume effect of the correlator, while no modelling of the contribution from the correlator tail is assumed.
In the future, we plan to further reduce the model-dependence employing, where available, a full lattice determination of $F_\pi(\omega)$~\cite{Andersen:2018mau,Erben:2019nmx} instead of the \ac{GS} parametrization.

The $F_\pi(\omega)$-based model provides a good spectral representation of the correlator up to the three-pion threshold, thus we use it for the correlator correction at times $x_0>\ti$, with $\ti=(M_\pi L/4)^2/M_\pi$.
At smaller times $x_0\leq\ti$ we compute the difference between the infinite- and finite-volume correlator in scalar QED (\emph{i.e.}\ NLO \textchi PT)~\cite{Francis:2013qna,DellaMorte:2017dyu}
\begin{equation}
  G^{33}(x_0,\infty) - G^{33}(x_0,L) = \frac{1}{3} \left( \int\frac{\dd[3]{\vec{k}}}{(2\cpi)^3} - \frac{1}{L^3}\sum_{\vec{k}} \right) \frac{\vec{k}^2+M_\pi^2}{\vec{k}^2} \e{-2x_0\sqrt{\vec{k}^2+M_\pi^2}} .
\end{equation}
This model omits the pion self-interaction and it is known to only account for a fraction of the finite-volume correction to $\SPi(Q^2)$ at $Q^2$ values of $\order{\si{\GeV\squared}}$~\cite{Aubin:2015rzx}.
However, comparing the correlator on two pairs of ensembles in Table~\ref{tab:ensemble} with the same parameters except for different physical volumes, we observe that the finite-volume correction at times $x_0\leq\ti$ is smaller than our uncertainties.
Therefore, we are confident that this rather simple short-time description is sufficient at the current level of precision, while recent developments~\cite{Hansen:2019rbh} show a promising path towards improving on it.

We can estimate the relative size of the finite-volume corrections.
For example, on ensembles D200, N302 and J303, the ratio of the correction over the total $\Dalphahad$ at $Q^2=\SI{1}{\GeV\squared}$ amounts to more than $\SI{1}{\percent}$.
The contribution, although small in absolute terms, is already bigger than the $\SI{0.5}{\percent}$ statistical uncertainty on these ensembles and therefore needs to be included.

\subsection{Signal-to-noise ratio}
\label{sec:SN}

Both the correlator and its variance can be expressed, using the spectral decomposition, as an infinite tower of exponentials, but with different decay rates and amplitudes.
In general, this leads to an exponential deterioration of the signal with respect to the noise with Euclidean time, known as the \acl{SN} problem.
As shown in Section~\ref{sec:TMR}, this problem has a larger impact on the statistical error of $\SPi(Q^2)$ at smaller $Q^2$.
Although it affects all flavour components, it is more acute for the light correlator, whose contribution remains sizeable longer in time.

As described in Ref.~\cite{DellaMorte:2017dyu}, one possibility is to model the tail of the correlator with a single exponential, \emph{i.e.}\
\begin{equation}
  G(x_0) = \begin{cases}
    \mathrm{data},		&	x_0 < \tcut , \\
    A\e{-m_V x_0},	&	x_0 \geq \tcut ,
  \end{cases}
\end{equation}
where $A$ and $m_V$ are obtained fitting the correlator.
In our case, due to the high statistics available, $\tcut$ lies between \SI{1.5}{\fm} and \SI{2.5}{\fm}, depending on the ensemble.
However, there is no guarantee that a single state describes the tail of the correlator.
This is particularly true for light ensembles in a large volume, where both the $\rho$ resonance and the tower of two-$\pi$ states are long range contributions.
Therefore, this method has its limitations, and we choose not to use it.
The results presented in the following sections are obtained using the correlator data until the last time slice available.
As explained in Section~\ref{sec:TMR}, the contribution from the tail to $\SPi(Q^2)$ is smaller than in the case of $a_\mu^{\mathrm{HLO}}$.

In the future, we are considering improving the statistical error on $\SPi(Q^2)$ at small $Q^2$, especially on light-pion ensembles, substituting the correlator at long times with a spectral reconstruction, and controlling the systematics \emph{e.g.}\ with the bounding method~\cite{Lehner:talkLGT16,Gerardin:2019rua}. 
In its simplest incarnation, this relies on the fact that $G(x_0)$ at $x_0\geq\tcut$ is bounded by
\begin{equation}
  0 \leq G(\tcut) \e{-E_{\mathrm{eff}}(\tcut)(x_0-\tcut)} \leq G(x_0) \leq G(\tcut) \e{-E_0(x_0-\tcut)} ,
\end{equation}
where $E_{\mathrm{eff}}(x_0) = -\dd\log(G(x_0))/\dd{x_0}$ is the effective mass and $E_0$ the ground state energy, to obtain the $\tcut$ so both bounds yield the same result for the observable.

\section{Numerical results}
\enlargethispage{1\baselineskip}

\begin{figure}[t]
  \centering
  \scalebox{.62}{\input{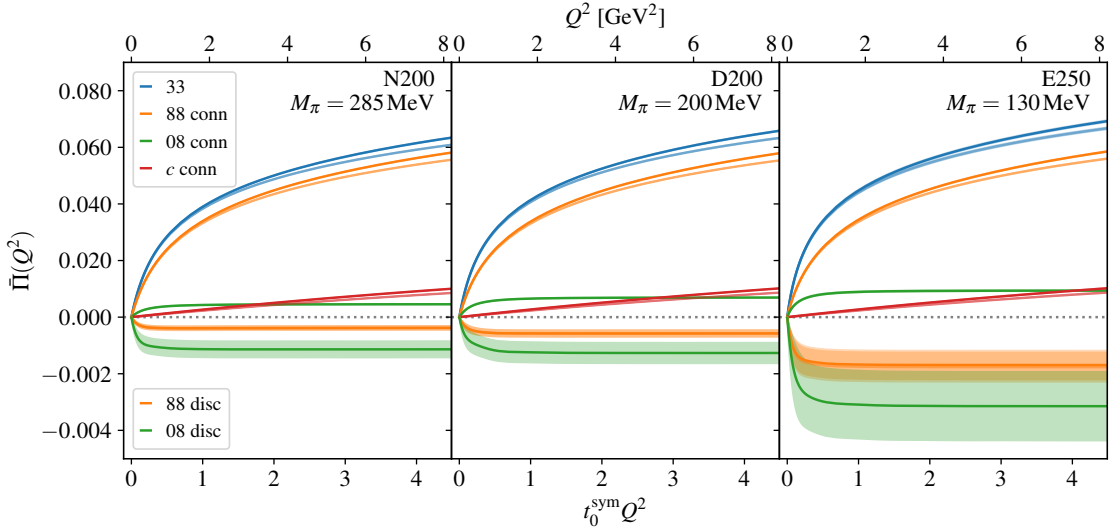}}
  \caption{%
    Running with energy $Q^2$ of different contributions to $\SPi(Q^2)$ on three different ensembles at $a\approx\SI{0.064}{\fm}$.
    The conserved, local discretization is shown and, when available, the local, local discretization in a lighter colour shade.
    The negative side of the vertical axis of the plot is inflated by a factor \num{10} with respect to the positive side.
  }\label{fig:running_lattice}
\end{figure}

\begin{table}[b]
  \caption{%
    Estimate of connected and disconnected contributions to $\num{e5}\cdot\SPi(\SI{1}{\GeV\squared})$ for the conserved-local (c.l.) and, when available, local-local (l.l.) discretizations, on three different ensembles at $a\approx\SI{0.064}{\fm}$.
  }\label{tab:results_lattice}
  \centering
  \begin{tabular}{rS[table-format=4(2)]S[table-format=4(2)]S[table-format=3(2)]S[table-format=3.1(2)]S[table-format=+3(2),table-sign-mantissa]S[table-format=+3(3),table-sign-mantissa]}
    \toprule
    {$\times\num{e-5}$} & {$\SPi^{33}$} & {$\SPi_{\mathrm{conn}}^{88}$} & {$\SPi_{\mathrm{conn}}^{08}$} & {$\SPi_{\mathrm{conn}}^c$} & {$\SPi_{\mathrm{disc}}^{88}$} & {$\SPi_{\mathrm{disc}}^{08}$} \\
    \midrule
    N200 c.l. &  3002(11) &  2537( 5) &  393( 6) &  266.2( 5) &  -39( 7) & -110(29) \\
         l.l. &  2962(11) &  2497( 5) &          &  228.4( 5) &  -39( 7) &          \\
    D200 c.l. &  3226(13) &  2526( 5) &  600( 8) &  270.3( 6) &  -57(12) & -120(36) \\
         l.l. &  3185(14) &  2485( 5) &          &  232.5( 5) &  -57(12) &          \\
    E250 c.l. &  3552(36) &  2594(12) &  826(21) &  271.8(40) & -164(50) & -301(120) \\
         l.l. &  3511(36) &  2553(12) &          &  233.7(34) & -171(51) &          \\
    \bottomrule
  \end{tabular}
\end{table}{}

Figure~\ref{fig:running_lattice} shows the running of different contributions to $\SPi(Q^2)$, defined through the correlators in Eq.~\eqref{eq:corr_flavour}, as a function of $Q^2$ on three different lattices at the same lattice spacing with increasingly lighter pions.
As one moves away from the $\SU(3)$-symmetric point, the $\SPi^{33}$ contribution increases while the $\SPi_{\mathrm{conn}}^{88}$ contribution decreases.
The (quenched) charm contribution is also shown to be relatively independent on the pion mass and linearly increasing in the range of $Q^2$ values.
The negative disconnected contributions are also shown, enlarged by a factor \num{10}.
They are obtained cutting the \ac{TMR} integration at $\tcut\approx\SI{2.5}{\fm}$, which results in a conservative statistical error.
Nevertheless, Table~\ref{tab:results_lattice} shows that the statistical error on the disconnected contribution is comparable to the connected one.
The exception is the E250 ensemble, with only a small number of disconnected loop measurements being available at the current time.
We will improve the disconnected contribution estimate by implementing the bounding method for the $I=0$ channel, in order to reduce the statistical error and correctly estimate the integration tail systematics.
It is worth noting that $\SPi_{\mathrm{disc}}^{88}(Q^2)$ is constant for $Q^2\gtrsim\SI{0.5}{\GeV\squared}$, as predicted by perturbation theory.

\subsection{Extrapolation to the physical point}
\label{sec:extrapolation}

\begin{table}[b]
  \caption{%
    Results extrapolated to the physical point for the different connected contributions to $\SPi$ for a range of $Q^2$ values.
    Only the connected contribution is included in $\Dalphahad$ and $\Dhad\sIIW$.
  }
  \label{tab:phys_results}
  \centering
  \begin{tabular}{S[table-format=1.1]S[table-format=1.5(2)]S[table-format=1.5(2)]S[table-format=1.5(2)]S[table-format=1.6(2)]S[table-format=+1.6(2),table-sign-mantissa]}
    \toprule
    {$Q^2$ [\si{\GeV\squared}]} & {$\SPi^{33}$} & {$\SPi_{\mathrm{conn}}^{88}$} & {$\SPi_{\mathrm{conn}}^{08}$} & {$\Dalphahad$} & {$\Dhad\sIIW$} \\
    \midrule
    0.1 & 0.00764(16) & 0.00456(8)  & 0.00267(8)  & 0.000841(16) & -0.000821(15) \\
    0.4 & 0.02068(26) & 0.01354(19) & 0.00619(12) & 0.002310(29) & -0.002300(29) \\
    1.0 & 0.03274(34) & 0.02340(27) & 0.00809(13) & 0.003718(38) & -0.003770(40) \\
    2.0 & 0.04242(38) & 0.03226(33) & 0.00880(13) & 0.004876(45) & -0.005012(47) \\
    3.0 & 0.04805(41) & 0.03768(36) & 0.00898(13) & 0.005558(48) & -0.005752(50) \\
    4.0 & 0.05202(42) & 0.04157(38) & 0.00904(13) & 0.006041(50) & -0.006278(53) \\
    5.0 & 0.05508(43) & 0.04461(39) & 0.00907(13) & 0.006415(51) & -0.006687(54) \\
    \bottomrule
  \end{tabular}
\end{table}

\begin{figure}[t]
  \centering
  \scalebox{.62}{\input{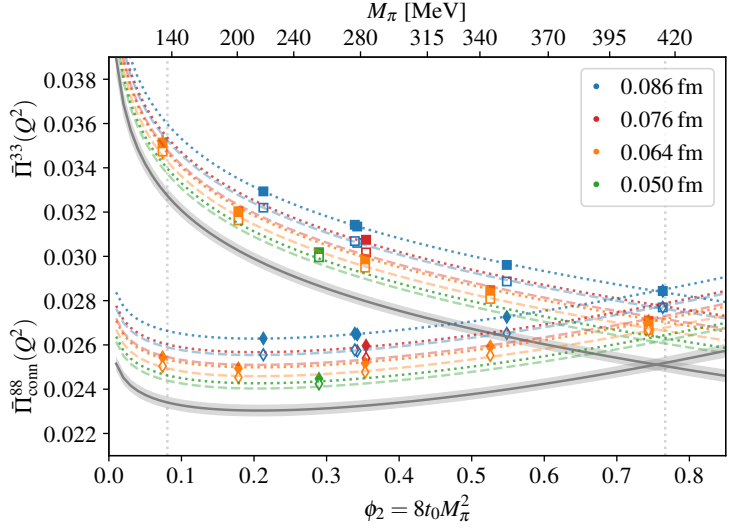}}
  \caption{%
    Combined extrapolation of $\SPi^{33}(Q^2)$ and $\SPi_{\mathrm{conn}}^{88}(Q^2)$ at $Q^2=\SI{1}{\GeV\squared}$ to the physical point.
    Filled symbols denote the conserved, local discretization, while open symbols denote the local, local one.
  }\label{fig:extrapolation}
\end{figure}

Four different lattice spacings and several quark masses, including the physical ones, allow us to extrapolate our results to the physical point.
In these proceedings, we consider only the extrapolation of the light and strange connected contributions, which is obtained with a combined fit of the two discretizations of $\SPi^{33}(Q^2)$ and $\SPi_{\mathrm{conn}}^{88}(Q^2)$.
At a given fixed $Q^2$, each \ac{HVP} function contribution in the combined fit is modelled with
\begin{multline}\label{eq:model}
  \SPi(a^2/t_0,\phi_2,\phi_4) = \SPi^{\mathrm{sym}} + \delta_2 a^2/t_0 + \delta_3 (a^2/t_0)^{3/2} + \beta_{21} a^2(\phi_2-\phi_2^{\mathrm{sym}})/t_0 \\
  + \gamma_1 (\phi_2-\phi_2^{\mathrm{sym}}) + \gamma_2 (\log\phi_2-\log\phi_2^{\mathrm{sym}}) + \gamma_4 (\phi_2-\phi_2^{\mathrm{sym}})^2 + \eta_1 (\phi_4-\phi_4^{\mathrm{sym}}) ,
\end{multline}
where $\phi_2=8\tnotsym M_\pi^2$ and $\phi_4=8\tnotsym(M_K^2+M_\pi^2/2)$ are proxies for the $M_\pi$ and $M_K$ dependence.
For the continuum limit extrapolation, $a^2$ and $a^3$ terms are included with different coefficients for the two different discretizations.
To fit the dependence on the meson masses, we considered different models.
In the fit shown in Figure~\ref{fig:extrapolation}, the $\SPi^{33}(Q^2)$ interpolation includes the $\gamma_1^{33}$ term $\sim M_\pi^2$ and the $\gamma_2^{33}$ term $\sim\log M_\pi^2$, while the $\SPi_{\mathrm{conn}}^{88}(Q^2)$ interpolation includes an independent $\gamma_1^{88}$ term and a $\gamma_4^{88}$ term $\sim M_\pi^4$, while we constrain $\gamma_2^{88}=\gamma_2^{33}/3$.
The latter term is included to match the diverging behaviour of $\SPi^{33}(Q^2)$ with $M_\pi\to 0$, that is present in $\SPi_{\mathrm{conn}}^{88}(Q^2)$, with a factor of one third, due to the missing disconnected contribution~\cite{Gerardin:2019rua}.

The combined fit includes the uncertainties in the determination of the pion and kaon masses, including the correlation with the \ac{HVP} function determination, that are obtained from a dedicated computation.
For every ensemble $l$, we build a residue vector
\begin{equation}
 \nu_l = \begin{pmatrix}
    \phi_2^l \\
    \SPi_{\mathrm{cl}}^{33}(a^2/\tnotsym,\phi_2^l,\phi_4^l) \\
    \SPi_{\mathrm{ll}}^{33}(a^2/\tnotsym,\phi_2^l,\phi_4^l) \\
    \phi_4^l \\
    \SPi_{\mathrm{cl}}^{88}(a^2/\tnotsym,\phi_2^l,\phi_4^l) \\
    \SPi_{\mathrm{ll}}^{88}(a^2/\tnotsym,\phi_2^l,\phi_4^l)
 \end{pmatrix} - \begin{pmatrix}
    8\tnotsym M_\pi^2 \\
    \SPi_{\mathrm{cl}}^{33} \\
    \SPi_{\mathrm{ll}}^{33} \\
    8\tnotsym(M_K^2+M_\pi^2/2) \\
    \SPi_{\mathrm{cl}}^{88} \\
    \SPi_{\mathrm{ll}}^{88}
 \end{pmatrix}_{\mathrm{data}}
\end{equation}
and we minimize the $\chi^2=\sum_{l\in\{\mathrm{ensembles}\}} \nu_l^\mathsf{T} C^{-1}_l \nu_l$, with the necessary adaptations for $\SU(3)$-symmetric ensembles where $\SPi^{33}$ and $\SPi^{88}$, and $M_\pi$ and $M_K$, are not independent.

This choice leads to an acceptable fit with an additional cut on the physical lattice volume that excludes lattices with $L<\SI{2.5}{\fm}$.
At $Q^2=\SI{1}{\GeV\squared}$, we have $\chi^2/\mathrm{dof}=52.13/38=1.37$, which corresponds to a $p$-values of \num{0.0631}, and parameters
\begin{equation}
\begin{gathered}
  \SPi^{\mathrm{sym}}=\num{0.02516(29)}, \mkern12mu \phi_2^{\mathrm{sym}}=2\phi_4^{\mathrm{sym}}/3=\num{0.754(5)} , \mkern12mu \eta_1=\num{-0.011(4)} , \\
  \delta_2^{\mathrm{cl}}=\num{0.0126(31)}, \mkern12mu \delta_2^{\mathrm{ll}}=\num{0.0106(31)} , \mkern12mu \delta_3^{\mathrm{cl}}=\num{-0.005(4)}, \mkern12mu \delta_3^{\mathrm{ll}}=\num{-0.006(4)} , \mkern12mu \beta_{21}=\num{0.0004(23)}, \\
  \gamma_1^{33}=\num{-0.0017(9)}, \mkern12mu \gamma_1^{88}=\num{0.00363(33)}, \mkern12mu \gamma_2^{33}=\gamma_2^{88}/3=\num{-0.00291(28)} , \mkern12mu \gamma_4^{88}=\num{0.0026(5)} .
\end{gathered}
\end{equation}
The extrapolation to $a=0$ and to physical $\pi^0$ and $K^0$ masses gives $\SPi^{33}=\num{0.03274(34)}$ and $\SPi_{\mathrm{conn}}^{88}=\num{0.02340(27)}$.
Moreover, $\SPi_{\mathrm{conn}}^{08}=\sqrt{3}(\SPi^{33}-\SPi_{\mathrm{conn}}^{88})/2=\num{0.00809(13)}$.
Applying Eqs~\eqref{eq:Dhad_alpha} and~\eqref{eq:Dhad_sIIW} allows us to compute the leading hadronic connected contribution from three-flavour \ac{QCD} to the running of $\alpha$ and $\sIIW$ at $Q^2=\SI{1}{\GeV\squared}$
\begin{equation}
   \Dalphahad(-\SI{1}{\GeV\squared}) = \num{0.003718(38)} , \qquad \Dhad\sIIW(-\SI{1}{\GeV\squared}) = \num{-0.003770(40)} .
\end{equation}

\subsection{The running with energy}

\begin{figure}[t]
  \centering
  \scalebox{.62}{\input{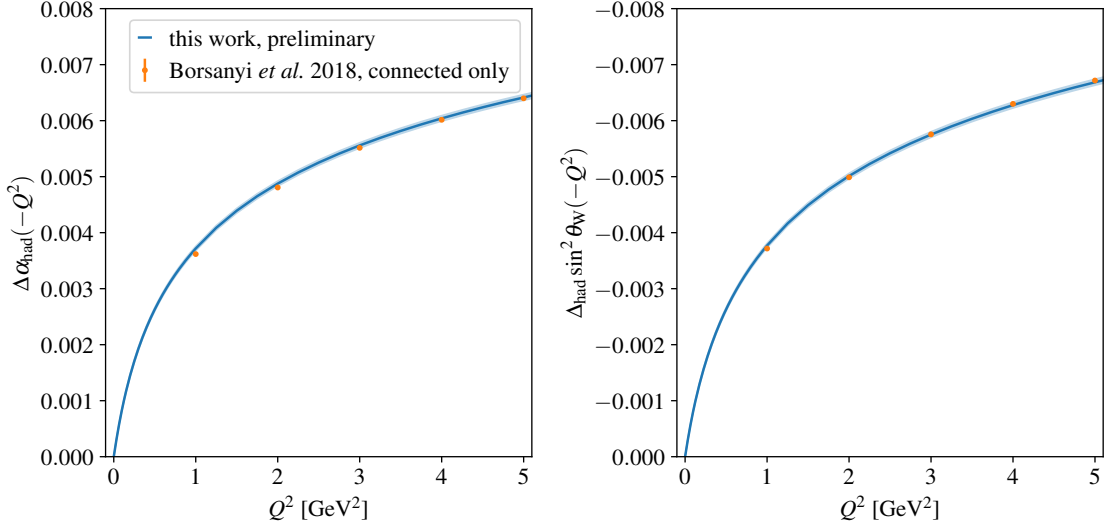}}
  \caption{%
    Leading hadronic (connected) contribution from three-flavour \ac{QCD} to the running of the electromagnetic coupling $\alpha$ (left) and the electroweak mixing angle $\sIIW$ (right) as a function of the space-like momentum transfer $Q^2$.
    The blue line (with error band) is the extrapolation of our lattice data, while the orange points are constructed from the \ac{HVP} values from Table~S3 of the supplemental material of Ref.~\cite{Borsanyi:2017zdw}.
  }\label{fig:running_phys}
\end{figure}

Performing the fit described in Section~\ref{sec:extrapolation} for different values of $Q^2$ results in a mild dependence of the $\chi^2$ on $Q^2$ up to around $\SI{2}{\GeV\squared}$.
At higher $Q^2$ discretization effects become the dominant systematics and the quality of the fit, using the current model, starts to deteriorate.
Results extrapolated to the physical point for a number of $Q^2$ values are given in Table~\ref{tab:phys_results}, and in Figure~\ref{fig:running_phys} we plot the \ac{HVP} contribution to the running of $\alpha$ and $\sIIW$.
In both plots, we compare our preliminary result to the corresponding contribution from the percent-level lattice determination of Ref.~\cite{Borsanyi:2017zdw}, given in the supplemental material for five $Q^2$ values.
The agreement is good, with a small tension only at the smallest $Q^2=\SI{1}{\GeV\squared}$.

\section{Conclusions and outlook}

We computed on the lattice the leading hadronic contribution to the running of the electromagnetic coupling $\alpha$ and of the electroweak mixing angle $\thetaW$.
After extrapolating to the physical point, the statistical error on the connected contribution amounts to $\approx\num{4e-5}$ which, for both quantities, is \SI{1}{\percent} of the contribution at $Q^2=\SI{1}{\GeV\squared}$.
We also presented results for the disconnected contribution, not yet extrapolated at the physical point, which show that the statistical error is also around \SI{1}{\percent} of the total contribution.
We included finite-volume corrections, that are instrumental for obtaining a reliable physical-point extrapolation.
The precision of our results is comparable to the phenomenological estimate, which is around \SI{0.7}{\percent} and \SI{1}{\percent} for the two quantities respectively~\cite{Jegerlehner:2017zsb}.
In particular, flavour contributions are naturally singled out on the lattice. 
Thus, our results can be combined with the phenomenological analysis of the electroweak mixing angle to reduce its flavour separation systematics.

We are currently working on increasing the statistics of disconnected loops at physical meson masses.
Subsequently, we will extrapolate the disconnected and the quenched charm contribution to the physical point.
Moreover, a full assessment of systematic errors is missing from the results in Table~\ref{tab:phys_results}.
We plan to assess the systematics introduced by the extrapolation by varying the choice of the fit function, and, in the future, adding a new ensemble at the fine lattice spacing $a\approx\SI{0.050}{\fm}$ with a lighter pion mass of $M_\pi\approx\SI{175}{\MeV}$.
We plan, as described in Section~\ref{sec:SN}, to improve the treatment of the long-time tail of the lighter ensembles with the bounding method.
The scale setting introduces an error that can be estimated as explained in Section~B.2 of Ref.~\cite{DellaMorte:2017dyu}.
Also, a small systematics from the mistuning of the charm hopping parameter needs to be added.
Finally, to compare with the physical world, isospin-breaking effects from non-degenerate $u$ and $d$ quark masses and QED are to be included~\cite{Risch:2018ozp,Risch:2019xio}.

{\noindent\footnotesize \textbf{Acknowledgements:}
We thank Jens Erler, Fred Jegerlehner and Daniel Mohler for valuable discussions.
Calculations for this project have been performed on the HPC clusters \enquote{clover} and \enquote{himster2} at Helmoltz-Institut Mainz and \enquote{Mogon II} at JGU Mainz, on the BG/Q system \enquote{JUQUEEN} at Jülich Supercomputing Center (JSC), and on \enquote{Hazel Hen} at Höchstleistungsrechenzentrum Stuttgart (HLRS).
The authors gratefully acknowledge the support of the Gauss Centre for Supercomputing  and the John von Neumann-Institut für Computing (NIC) for project HMZ21 at JSC and project GCS-HQCD at HLRS.
Our programs use the deflated SAP + GCR solver from the \textsc{openQCD} package, as well as the QDP++ library.
We are grateful to our colleagues in the \ac{CLS} initiative for sharing ensembles.}

\bibliographystyle{JHEP}
\bibliography{./biblio.bib}

\end{document}